\documentclass[useAMS]{mn2e}
\usepackage{graphicx}
\usepackage{epsfig}

\newcommand{\hii}{H{\sc ii}}

\def\vhel{\ifmmode{V_{{\rm HEL}}}\else{$V_{{\rm HEL}}$}\fi}
\def\vsys{\ifmmode{V_{\rm sys}}\else{$V_{\rm sys}$}\fi}
\def\kms{\ifmmode{~{\rm km\,s}^{-1}}\else{~km~s$^{-1}$}\fi}
\def\vlsr{\ifmmode{v_{\rm lsr}}\else{$v_{\rm lsr}$}\fi}

\def\ltsim{\ifmmode\stackrel{<}{_{\sim}}\else$\stackrel{<}{_{\sim}}$\fi}
\def\gtsim{\ifmmode\stackrel{>}{_{\sim}}\else$\stackrel{>}{_{\sim}}$\fi}

\def\reff@jnl#1{{\rm#1\/}}

\def\aj{\reff@jnl{AJ}}                  % Astronomical Journal
\def\araa{\reff@jnl{ARA\&A}}            % Annual Review of Astron and Astrophys
\def\apj{\reff@jnl{ApJ}}                % Astrophysical Journal
\def\apjl{\reff@jnl{ApJ}}               % Astrophysical Journal, Letters
\def\aplett{\reff@jnl{ApJ}}             % COPY OF ABOVE FOR ADS aplett command
\def\apjs{\reff@jnl{ApJS}}              % Astrophysical Journal, Supplement
\def\ao{\reff@jnl{Appl.Optics}}         % Applied Optics
\def\apss{\reff@jnl{Ap\&SS}}            % Astrophysics and Space Science
\def\aap{\reff@jnl{A\&A}}               % Astronomy and Astrophysics
\def\aapr{\reff@jnl{A\&A~Rev.}}         % Astronomy and Astrophysics Reviews
\def\aaps{\reff@jnl{A\&AS}}             % Astronomy and Astrophysics, Supplement
\def\azh{\reff@jnl{AZh}}                        % Astronomicheskii Zhurnal
\def\baas{\reff@jnl{BAAS}}              % Bulletin of the AAS
\def\jrasc{\reff@jnl{JRASC}}            % Journal of the RAS of Canada
\def\memras{\reff@jnl{MmRAS}}           % Memoirs of the RAS
\def\mnras{\reff@jnl{MNRAS}}            % Monthly Notices of the RAS
\def\pra{\reff@jnl{Phys.Rev.A}}         % Physical Review A: General Physics
\def\prb{\reff@jnl{Phys.Rev.B}}         % Physical Review B: Solid State
\def\prc{\reff@jnl{Phys.Rev.C}}         % Physical Review C
\def\prd{\reff@jnl{Phys.Rev.D}}         % Physical Review D
\def\prl{\reff@jnl{Phys.Rev.Lett}}      % Physical Review Letters
\def\pasp{\reff@jnl{PASP}}              % Publications of the ASP
\def\pasj{\reff@jnl{PASJ}}              % Publications of the ASJ
\def\qjras{\reff@jnl{QJRAS}}            % Quarterly Journal of the RAS
\def\skytel{\reff@jnl{S\&T}}            % Sky and Telescope
\def\solphys{\reff@jnl{Solar~Phys.}}    % Solar Physics
\def\sovast{\reff@jnl{Soviet~Ast.}}     % Soviet Astronomy
 \def\ssr{\reff@jnl{Space~Sci.Rev.}}     % Space Science Reviews
\def\zap{\reff@jnl{ZAp}}                        % Zeitschrift fuer Astrophysik
\def\nat{\reff@jnl{Nature}}             % Nature 

\def\LaTeX{L\kern-.36em\raise.3ex\hbox{a}\kern-.15em
    T\kern-.1667em\lower.7ex\hbox{E}\kern-.125emX}

\def\deg{^\circ}
\def\h2{{\rm H}{\sc ii}}

\begin{document}

\title[CBI limits on 31~GHz excess emission in southern H{\sc ii} regions]{CBI limits on 31~GHz excess emission in southern H{\sc ii} regions}

\author[C. Dickinson et al.]{C. Dickinson,$^{1,2}$\thanks{E-mail: Clive.Dickinson@jpl.nasa.gov} R.D. Davies,$^{3}$ L. Bronfman,$^{4}$ S. Casassus,$^{4}$ R.J. Davis,$^{3}$
\newauthor T.J. Pearson,$^{2}$ A.C.S. Readhead$^{2}$ and P.N. Wilkinson$^{3}$ \\
$^1$Jet Propulsion Laboratory, 4800 Oak Grove Drive, M/S 169-327, Pasadena, CA 91109, U.S.A. \\ 
$^2$Chajnantor Observatory, California Institute of Technology, 1200 E. California Blvd., M/S 105-24, Pasadena, CA 91125, U.S.A \\ 
$^3$Jodrell Bank Observatory, University of Manchester, Macclesfield, Cheshire, SK11 9DL.\\
$^4$Departamento de Astronom{\'\i}a, Universidad de Chile,  Casilla 36-D, Santiago, Chile.}
%%%%%%%%%%%%%%%%%%%%%%%%%%%%%%%%%%%%%%%%%%%%%%%%%%%%%%%%%%%%%%%%%%%%%%%%

\date{Received **insert**; Accepted **insert**}
       
\pagerange{\pageref{firstpage}--\pageref{lastpage}} 
\pubyear{}

\maketitle
\label{firstpage}

%%%%%%%%%%%%%%%%%%%%%%%%%%%%%%%%%%%%%%%%%%%%%%%%%%%%%%%%%%%%%%%%%%%%%%%%%

\begin{abstract}

We have mapped four regions of the southern Galactic plane at 31~GHz
with the Cosmic Background Imager. From the maps, we have extracted
the flux densities for six of the brightest \hii~regions in the
southern sky and compared them with multi-frequency data from the
literature. The fitted spectral index for each source was found to be
close to the theoretical value expected for optically thin free-free
emission, thus confirming that the majority of flux at 31~GHz is due
to free-free emission from ionised gas with an electron temperature of
$\approx 7000-8000$~K.

We also found that, for all six sources, the 31~GHz flux density was
slightly higher than the predicted value from data in the
literature. This excess emission could be due to spinning dust or
another emission mechanism. Comparisons with $100~\mu$m data indicate
an average dust emissivity of $3.3\pm1.7~\mu$K~(MJy/sr)$^{-1}$, or a
95 per cent confidence limit of $<6.1~\mu$K~(MJy/sr)$^{-1}$. This is
lower than that found in diffuse clouds at high Galactic latitudes by
a factor of $\sim 3-4$. The most significant detection ($3.3\sigma$)
was found in $G284.3-0.3$ (RCW49) and may account for up to $\approx
30$ per cent of the total flux density observed at 31~GHz. Here, the
dust emissivity of the excess emission is
$13.6\pm4.2~\mu$K~(MJy/sr)$^{-1}$ and is within the range observed at
high Galactic latitudes.

Low level polarised emission was observed in all six sources with
polarisation fractions in the range $0.3-0.6$~per cent. This is likely
to be mainly due to instrumental leakage and is therefore upper an
upper limit to the free-free polarisation. It corresponds to an upper
limit of $\sim1$ per cent for the polarisation of anomalous emission.

\end{abstract}


%%%%%%%%%%%%%%%%%%%%%%%%%%%%%%%%%%%%%%%%%%%%%%%%%%%%%%%%%%%%%%%%%%%%%%%%%%
%%%%%%%%%%%%%%%%%%%%%%%%%%%%%%%%%%%%%%%%%%%%%%%%%%%%%%%%%%%%%%%%%%%%%%%%%%

\setcounter{figure}{0}

\section{INTRODUCTION}
\label{sec:introduction}

The characterisation of diffuse emission from the Galaxy is important
for the detailed study of fluctuations in the cosmic microwave
background (CMB). Knowledge of the spectral shape and spatial
morphology is important for the accurate subtraction of the foreground
emission. This in turn provides a more precise view of the CMB
anisotropies thus providing the most reliable cosmological
information. The frequency range of greatest interest for CMB
observations is $\sim 30-200$~GHz, which is close to the minimum of
foreground emission at $\sim 70$~GHz (Bennett et al.~2003b). The
diffuse Galactic foregrounds include synchrotron, free-free and
vibrational (thermal) dust emissions. However there is considerable
evidence, from deep CMB data at high Galactic latitudes, for an
additional component that emits in the frequency range $\sim
10-60$~GHz (Kogut et al. 1996; Leitch et al.~1997; de Oliveira-Costa
et al.~2002,2004; Banday et al.~2003; Finkbeiner, Langston \& Minter,
2004; Lagache 2003; Davies et al.~2006; Fern{\'a}ndez-Cerezo et
al.~2006). The data show a strong correlation with FIR ($\sim
100~\mu$m) emission that suggests a connection to dust grains. The
most popular candidate for this anomalous component is rapidly
spinning small dust grains (Draine \& Lazarian 1998a,b), referred to
as spinning dust. Models of spinning dust emission predict a strongly
peaked spectrum at $~20-40$~GHz, which appears to be broadly
consistent with the data. Furthermore, targetted observations of
Galactic sources have shown excess emission in this frequency range
(Finkbeiner et al. 2002,2004; Casassus et al.~2004,2006; Watson et
al. 2005; Scaife et al.~2007).

At radio/microwave frequencies ($\ltsim 100$~GHz), \h2 regions are
dominated by free-free (thermal bremsstrahlung) emission from ionised
gas with electron temperatures, $T_{e}\approx 8000$~K. The spectrum of
free-free radiation is well-understood; in the optically thin regime
($\gtsim 1~$GHz), it has a well-defined flux density spectral
index\footnote{Throughout the paper, the flux density spectral index,
$\alpha$, is defined as $S \propto \nu^{\alpha}$.} $\alpha\approx-0.1$
that does not vary greatly with frequency or $T_e$
\cite{Dickinson03}. However, emission from spinning dust could be
inherent since ion collisions with grains are expected to be the
largest contributory factor in maintaining large rotational velocities
required to produce significant rotational dust emission (Draine \&
Lazarian 1998b). Indeed, several detections are associated with \h2
regions (Watson et al.~2005) or PNe (Casassus et al. 2004,2007). One
previous tentative detection of spinning dust from the \h2 region
LPH96[201.663+1.643], which appeared to show a rising spectrum from
$5-10$~GHz, suggestive of spinning dust (Finkbeiner et al.~2002;
Finkbeiner~2004), was shown to be a spurious result with only little
room for spinning dust (Dickinson et al.~2006; Scaife et al.~2007).

In this paper we have imaged several southern \h2 regions with the
Cosmic Background Imager (CBI) to look for excess emission at 31~GHz,
which could be attributed to spinning dust. These are among the
brightest \h2 regions in the sky and are known to be dominated
by free-free emission at radio frequencies up to $\sim
100$~GHz. Furthermore, they contain copious amounts of dust within the
same volume, as traced by correlated FIR ($\sim 100~\mu$m)
emission. Using data from the literature, we model the spectrum of
free-free radiation and compare the CBI data to measure or place
limits on possible excess emission at 31~GHz. We also use CBI
polarisation data to measure and place upper limits on the
polarisation of free-free emission.

%%%%%%%%%%%%%%%%%%%%%%%%%%%%%%%%%%%%%%%%%%%%%%%%%%%%%%%%%%%%%%%%%%%%%%%%
%%%%%%%%%%%%%%%%%%%%%%%%%%%%%%%%%%%%%%%%%%%%%%%%%%%%%%%%%%%%%%%%%%%%%%%%

% Observation table to go on the 2nd page.
\begin{table*}
\caption{Summary of CBI observations.$^{*}$Integration times take into account data lost due to bad weather and data editing.}
\begin{tabular}{lcccl} \hline
Object/        &Centre R.A.         &Centre Dec.          &Integration$^{*}$ (hr)/ &Notes   \\  
Region         &(J2000)             &(J2000)              &noise level (Jy)&        \\ 
\hline
$G267.9-1.1$  &$08^{h}58^{m}55^{s}$&$-47\deg30^{m}58^{s}$ &1.25/       &Contains $G267.9-1.1$ (RCW38) and fainter extended components   \\
(RCW38)       &                    &                      &0.17        &to the north ($G267.8-0.9$) and to the east ($G268.1-1.0$).             \\

$G284.3-0.3$  &$10^{h}24^{m}20^{s}$&$-57\deg44^{m}57^{s}$ &0.43/       &Low level extension to the east. \\ 
(RCW49)       &                    &                      &0.14        &           \\

$G287.4-0.6$  &$10^{h}43^{m}52^{s}$&$-59\deg34^{m}33^{s}$ &2.28/       &Carina nebula (RCW53,NGC3372) covers a region $\sim 2\times2$ degrees. \\
(Carina nebula) &                  &                      &0.15        &2 bright spots: Car-I \& Car-II. \\

$G291.6-0.5$  &$11^{h}15^{m}00^{s}$&$-61\deg16^{m}00^{s}$ &1.15/       &Two distinct \h2 regions: $G291.6-0.5$ (NGC3603) \& \\ 
(RCW57)       &                    &                      &0.18        &$G291.3-0.7$ (NGC3576,RCW57)  \\
\hline
\end{tabular}
\label{tab:obs}
\end{table*}

\section{Data}

\subsection{The CBI interferometer}

The CBI is a 13-element interferometer, located on the high altitude
site (5080-m elevation) of Chajnantor Observatory, Chile. The 0.9-m diameter
Cassegrain antennas are co-mounted on a 6-m platform which tracks the
sky on a Alt-Az mount but at a constant parallactic angle through
rotation of the ``deck'' platform. This gives a static $u,v$-coverage
for a given configuration and given ``deck'' angle. Using the best
low-noise amplifiers  provides a typical system temperature $T_{\rm
sys} \sim 30$~K in a 10~GHz band centred at 31-GHz. The bandwidth is
split into 10 separate 1-GHz-wide bands from 26 to 36 GHz which can be
used to provide spectral information. Each antenna can measure a
single left (L) or right (R) circular polarisation mode therefore
allowing observations in total intensity (LL or RR) or polarisation
(LR or RL). Rotation of the deck allows the ``filling'' of the $u,v$
plane to improve beam quality in Stokes $I$ (total intensity) and
combinations of LR or RL thus allowing mapping of Stokes $Q$ and
$U$. We use the CBI in a compact configuration that gives optimal
sensitivity to extended objects and many redundant baselines. Baseline
lengths range from 1-m to 4-m, which corresponds to angular scales
$\sim 6$~arcmin to $\sim 24$ arcmin and  a primary beam FWHM of
$45.2$~arcmin at 31~GHz.

%%%%%%%%%%%%%%%%%%%%%%%%%%%%%%%%%%%%%%%%%%%%%%%%%%%%%%%%%%%%%%%%%%%%%%%%

\subsection{Observations}

Observations of several of the brightest southern \h2 regions were
made with the CBI in a compact configuration during the period April -
July 2005. Here we present data for four regions: $G267.9-1.1$
(RCW38), $G284.3-0.3$ (RCW49), $G287.4-0.6$ (Carina nebula) and
$G291.6-0.5/G291.3-0.7$ (RCW57). These were chosen for their
brightness, well-studied radio spectra and strong FIR emission that is
aligned with the radio emission. Each region was observed for $\sim
2-3$ hours, at a range of deck angles to improve $u,v$ coverage. Only short observations were required
since the images are limited by dynamic range due to beam
deconvolution residuals and calibration errors, that will dominate
over the thermal noise for the brightest sources. Longer integrations
would not significantly improve the signal-to-noise ratio, except for
allowing more deck rotations. A summary
of the observations is given in Table~\ref{tab:obs}. The noise level
was estimated from areas of the map well outside the primary beam.

%%%%%%%%%%%%%%%%%%%%%%%%%%%%%%%%%%%%%%%%%%%%%%%%%%%%%%%%%%%%%%%%%%%%%%%%

\subsection{Data reduction}

The data were reduced and calibrated using in-house software, {\sc
cbical}, originally written for CMB data
analysis (see Readhead et al.~2004a,b and references therein for more
details). The majority of data editing and flagging was done
automatically, such as removing bad antennas, baselines or channels
that were noisy or not working correctly.

Flux calibration was achieved by observations of bright calibrator
sources (primarily Jupiter) tied to the temperature of Jupiter of
$T_{J}=147.3\pm 1.8~K$ at 32~GHz (Readhead et al. 2004a). This, in
principle, gives an accuracy of 1.3 per cent. We note that
short-term gain variations and elevation corrections have not been
applied due to instabilities of the noise calibration diodes in the
CBI system. Checks on the data, by comparing flux densities on
different nights, showed these to be below the 1~per cent level.

Ground spillover is a source of relatively strong contamination at the
level of $\sim 0.5$~Jy on the shortest baselines of the CBI. Due to
the co-mounted design, filtering based on varying fringe rates of the
astronomical signal (e.g. Watson et al.~2003) cannot be used. For CMB
measurements, lead/trail fields or other strategies must be employed
for subtraction of ground spillover  (e.g. Pearson et
al.~2003). Fortunately,  for very bright objects ($\gtsim 10$~Jy), the
ground signal is essentially negligible. For the southern Galactic
plane (longitudes $\sim250-300$), lead/trail fields are difficult to
observe since the plane is approximately parallel to lines of constant
declination, thus we have not performed the ground subtraction
technique. We find that the majority of the data are unaffected by
such contamination, which would be highly visible on the shortest
(1-m) baselines.

\subsection{Imaging and fitting}
\label{sec:imaging}

Imaging of the visibility data was carried out using the {\sc difmap}
package employing uniform weights to give optimal resolution since we
are mainly limited by dynamic range (typically 500:1 for the CBI)
rather than thermal noise. The dirty images were deconvolved using the
CLEAN algorithm \cite{Hogbom74}. Primary beam corrections were applied
to the CLEAN components directly so that each of the frequency
channels were corrected separately with a Gaussian function of FWHM
$45.2 \times (\nu/31~{\rm GHz})$ arcmin, which is a good approximation
to the measured CBI primary beam (Pearson et al.~2003).

The incomplete $u,v$ coverage of interferometric data can potentially
lead to loss of flux for sources that are extended relative to the
synthesised beam; $\sim 6$~arcmin for these data. The exact amount of
flux loss depends on the structure of the source and the $u,v$
coverage. To estimate the flux loss for CBI maps presented in this
paper, we simulated observations based on $100~\mu$m
{\it IRIS}\footnote{Throughout the paper we use the recently re-processed
{\it IRAS} 100~$\mu$m map of Miville-Desch{\^e}nes \& Lagache~(2005), ``{\it IRIS}'', which retains optimal
resolution (4.3~arcmin) while removing the majority of artifacts such
as striping.} maps using the same CBI real visibilities to define the
$u,v$-coverage for each region. This is particularly important for
complex extended structures. However, we found that the peak and
integrated flux densities for the individual sources studied were
within 5 per cent of the real values\footnote{All fits were made using
the {\sc aips} task {\sc jmfit}, which provides integrated flux
densities and errors for the fitted parameters.}. For example,
$G267.9-1.1$, which is located in a region of extended emission, was
reduced by just 3 per cent. On the other hand, the integrated flux
density within a 30 arcmin radius was 58 per cent indicating that the
extended emission is more strongly affected. This shows that for
sources with angular diameters comparable to the synthesised beam,
($\sim 6$~arcmin or smaller), the fitted flux densities are not
significantly affected by the missing spatial frequencies and are
correct to better than a few per cent. We therefore make no flux loss
corrections for the relatively compact sources considered here. Such
corrections would only increase the CBI 31~GHz flux densities quoted
in Table~\ref{tab:flux}.

To estimate a possible 31~GHz excess, data at lower frequencies were
used to make a power-law fit of the form $S=S_{31}(\nu_{\rm
GHz}/31)^{\alpha}$. We used data that were believed to be reliable and
did not include very low frequencies ($\ltsim 1$~GHz) where the
free-free emission becomes optically thick and the spectrum no longer
obeys a simple power-law. The best-fitting power-law was then used to
estimate the flux density at 31~GHz, which was compared to the
observed value from the CBI. Fitted flux densities include an error
term due to the fitting procedure. However, we include an additional
error of 2 per cent for instabilities in fitting Gaussians when
choosing different box sizes. In general, we found the fits to be
relatively stable to changes in box size. Experimentation with
sensible choices of boxes showed that the integrated flux
densities could vary by $\sim 1-2$ per cent. The errors quoted for the
CBI fluxes therefore contain 3 components: an absolute calibration
error (1.3 per cent), a variable fitting error, and an additional 2
per cent error. This results in a typical CBI flux density error of
$\approx 5$ per cent.

%%%%%%%%%%%%%%%%%%%%%%%%%%%%%%%%%%%%%%%%%%%%%%%%%%%%%%%%%%%%%%%%%%
%%%%%%%%%%%%%%%%%%%%%%%%%%%%%%%%%%%%%%%%%%%%%%%%%%%%%%%%%%%%%%%%%%

\section{Results}

\subsection{$G267.9-1.1$ (RCW38)}

\begin{figure}
\begin{center}
\includegraphics[width=0.48\textwidth,angle=0]{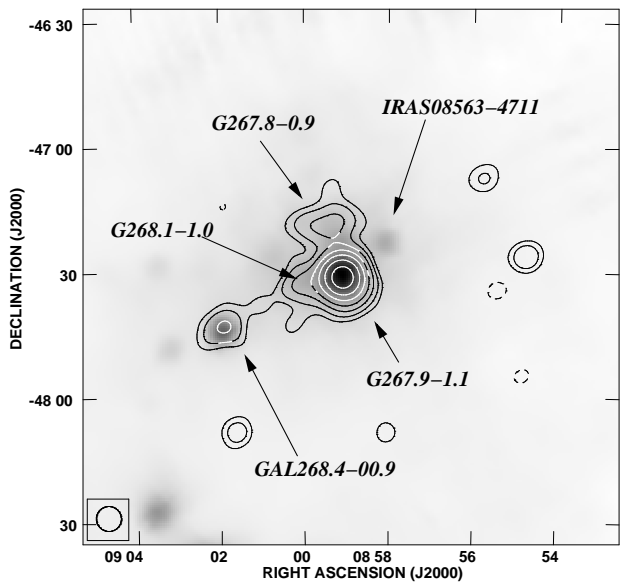}
 \caption{Map of the $G267.9-1.1$ (RCW38) region. CBI 31~GHz contours are overlaid on a greyscale image of the {\it IRAS} $100~\mu$m map, with a square-root stretch. Contours are at $-1~(dashed),1,2,4,8,16,32$ and 64 per cent of the peak intensity, $S_{p}=124.4$~Jy~beam$^{-1}$. The uniform-weighted beam is $6.0 \times 6.0$ arcmin. \label{fig:g267_cbi_map}}
\end{center}
\end{figure}

\begin{figure}
\begin{center}
\includegraphics[width=0.48\textwidth,angle=0]{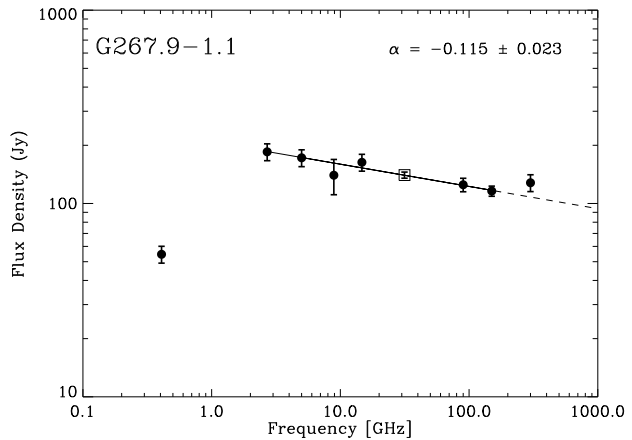}
 \caption{Spectrum of $G267.9-1.1$ (RCW38). Data points ({\it solid circles}) are integrated flux densities taken from the literature (see text). Uncertainties of 10 per cent were assumed when no error was given. The CBI 31~GHz value is plotted as a {\it square} symbol. The best fitting power-law ({\it solid line}) was fitted to the data over the range $5-150$~GHz and extended to higher frequencies ({\it dashed line}).}
\label{fig:g267_spec}
\end{center}
\end{figure}

$G267.9-1.1$, (also known as RCW38, Kes5; RA(J2000)$=08^{h}58^m55^s$,
Dec.(J2000)$=-47^d30^m58^s$) is one of the brightest and most dense
\hii~regions in the southern sky. The CBI 31~GHz
primary-beam-corrected map is shown in Fig.~\ref{fig:g267_cbi_map} as
contours overlaid on the {\it IRAS} 100~$\mu$m map. The peak 31~GHz
flux density is $124.4$~Jy~beam$^{-1}$ with a uniform-weighted
synthesised beam FWHM of $6.0$~arcmin. The dominant central component
is $G267.9-1.1$.  It has a fainter companion, $G267.8-0.9$ to the
north and also $G268.1-1.0$ to the east that can be seen as extensions
to the much brighter central object\footnote{The name $G268.0-1.0$ is
sometimes used in the older literature and usually refers to the
integrated emission from the entire region.}. The morphology is very
similar to the low frequency (0.4 and 5~GHz) maps of
\cite{Shaver70a,Goss70}.

The {\it IRAS} 100~$\mu$m emission is similar to that seen in the
radio (Fig.~\ref{fig:g267_cbi_map}). A compact source $\approx
13$~arcmin to the north-west of $G267.9-1.1$ is visible in the
$100~\mu$m image but is not seen in the CBI or other radio maps. It is
likely to be the IR source IRAS08563-4711 and appears to be associated
with the reflection nebula BRAN213. A relatively strong 100~$\mu$m
source is visible $\sim 30$~arcmin to the south-east of the RCW38
region and is also detected in the CBI map, but is significantly
attenuated by the 45~arcmin FWHM primary beam. This is identified as
the \hii~region GAL268.4-00.9 (IRAS09002-4732).

Three Gaussian components were found to be a very good fit to the
central region of the  primary-beam-corrected image. However, previous
data in the literature have been simply fitted with a single Gaussian
(plus a baseline offset) to each source so we have tried to replicate
this Gaussian fitting procedure to make a fairer comparison. This makes
a  difference of $\sim 5-10$ per cent in the integrated flux densities
due to the overlapping Gaussian contributions of several closely
spaced sources. The bright compact component ($G267.9-1.1$) contains
an integrated flux density, $S_{i}^{31}= 140.3 \pm 5.1$~Jy with a
deconvolved size of FWHM $2.6 \times 2.2$~arcmin. The northern
component ($G267.8-0.9$) has $S_i\approx 19$~Jy and is $\sim10 \times
7$~arcmin. A more extended Gaussian accounts for the eastern extension
with $S_i \approx 39$~Jy and is $\sim 15 \times 6$~arcmin but this was
not included when fitting for $G267.9-1.1$.

The spectrum of $G267.9-1.1$ is plotted in
Fig.~\ref{fig:g267_spec}. Data from the literature are plotted if they
were believed to be reliable in relation to the fitting procedure used
to determine the CBI flux densities. For example, the integrated flux
densities from VLA data at 1.4~GHz could not be reliably summed due to
significant flux losses which would not make a valid
comparison. Similarly, the lower resolution {\it WMAP} data (Bennett
et al.~2003a) is sensitive to local extended emissions. The data
points are at 0.4 and 5~GHz \cite{Shaver70b}, 2.7~GHz \cite{Day72},
8.9~GHz \cite{McGee75}, 14.7~GHz \cite{McGee81}, 90 and
150~GHz\footnote{These data were verified to be the most up-to-date
measurements at this frequency (P. Mauskopf; private communication).}
\cite{Coble03} and 300~GHz \cite{Cheung80}. The free-free emission is
clearly optically thick at 408~MHz and turns over at $\sim
1$~GHz. Above 5~GHz, the emission appears to be optically thin and is
best fitted by a power-law over the range $5-150$~GHz (omitting the
CBI data point) with flux density spectral index,
$\alpha=-0.115\pm0.023$. This agrees well with the theoretical value
of $\alpha\approx-0.12$ \cite{Dickinson03} for $\nu\sim30$~GHz and
$T_{e}\approx 7500$~K \cite{McGee75,Caswell87}. From
visibility-visibility correlations with the Parkes 6~cm map
\cite{Haynes78}, we found a consistent value of
$\alpha=-0.06\pm0.10$. Within the CBI band ($26-36$~GHz), the
best-fitting index was $\alpha=-0.15\pm0.09$, where the error was
estimated assuming a 2 per cent error over the range $26-36$~GHz. All
the data, including the CBI data point at 31~GHz, fit extremely well
with this simple free-free model, with a predicted 31~GHz flux density
$S_i^{31}=140.2\pm5.1$~Jy. Contributions from vibrational dust
emission are only important above 200~GHz. The fitted values to
$G267.9-1.1$ observations are summarised in Table~\ref{tab:flux}. The
data give an upper limit to a possible excess component of 14.2~Jy at
the 95 per cent confidence level (c.l.)\footnote{Throughout the paper,
upper limits are quoted at the 95 per cent confidence level (c.l.),
which is $\approx 2\sigma$.}. When the spectral index was fixed at
$\alpha=-0.12$, the upper limit remained at $<14.2$~Jy.

The FIR peak is well aligned in position with the peak in the
radio. We used the {\it IRAS} $100~\mu$m map to place limits on the
relative dust emission. Assuming a dust emissivity of
$10~\mu$K~(MJy/sr)$^{-1}$ at 31~GHz, the CBI-simulated $100~\mu$m map
results in a peak flux density of 13.8~Jy~beam$^{-1}$ and an
integrated flux density of 15.5~Jy. This corresponds to an upper limit
on the dust emissivity of $<9.2~\mu$K~(MJy/sr)$^{-1}$; see
Table~\ref{tab:flux}.

\begin{table*}
\caption{31~GHz integrated flux densities and derived limits on 31~GHz excess emission. Errors are quoted at the $1\sigma$ level while upper limits are given at 95 per cent ($\approx 2\sigma$) confidence level. Fits were made for both a floating and fixed spectral index. $^{*}$For $G291.3-0.7$, the CBI 31~GHz data point was included when the spectral index was fitted for. FWHM is the deconvolved size.}
\begin{tabular}{lcccccc} \hline
Source &Fitted &FWHM &Spectral index $\alpha$&Predicted &31~GHz excess &Excess $100~\mu$m emissivity\\  
   &$S_{i}^{31}$ (Jy) &(arcmin) &$(S\propto \nu^{\alpha})$ &$S_{i}^{31}$ (Jy) &(Jy) &[$\mu$K~(MJy/sr)$^{-1}$]     \\ 
\hline
$G267.9-1.1$ &$140.3\pm5.1$&$2.6\times2.2$&$-0.115\pm0.023$&$140.2\pm5.1$&$<14.2$&$<9.2$ \\ 
             &             &              &$-0.12$         &$140.3\pm5.1$&$<14.2$&$<9.2$ \\

$G284.3-0.3$ &$146.5\pm5.2$&$7.8\times5.6$&$-0.220\pm0.074$&$99.6\pm13.4$&$46.9\pm14.4$&$13.6\pm4.2$ \\ 
             &             &              &$-0.12$         &$117.0\pm5.7$&$29.5\pm7.8$ &$8.6\pm2.3$  \\

Car-I        &$83.9\pm7.6$&$8.8\times6.1$&$-0.145\pm0.038$&$79.8\pm8.3$ &$<24.8$&$<6.1$  \\ 
             &             &              &$-0.12$        &$84.8\pm3.8$ &$<16.0$&$<5.7$  \\

Car-II       &$92.1\pm8.1$&$9.4\times6.9$&$-0.101\pm0.048$&$77.5\pm11.1$&$<38.1$     &$<15.9$ \\
             &             &              &$-0.12$        &$73.4\pm3.7$ &$18.7\pm8.9$&$7.8\pm3.7$ \\

$G291.6-0.5$ &$158.7\pm5.8$&$7.1\times7.1$&$-0.071\pm0.078$&$143.2\pm19.0$&$<50.3$     &$<15.7$ \\ 
             &             &              &$-0.12$         &$132.5\pm6.6$ &$26.2\pm8.8$&$12.3\pm4.3$\\

$G291.3-0.7^{*}$ &$88.8\pm3.3$&$2.6\times2.6$&$-0.161\pm0.006$&$88.6\pm0.3$&$<6.6$ &$<6.7$   \\ 
             &             &              &$-0.12$        &$85.5\pm6.1$&$<15.8$    &$<16.1$   \\
\hline
\end{tabular}
\label{tab:flux}
\end{table*}

%%%%%%%%%%%%%%%%%%%%%%%%%%%%%%%%%%%%%%%%%%%%%%%%%%%%%%%%%%%%%%%%%%

\subsection{$G284.3-0.3$ (RCW49)}
\label{sec:g284}

\begin{figure}
\begin{center}
\includegraphics[width=0.48\textwidth,angle=0]{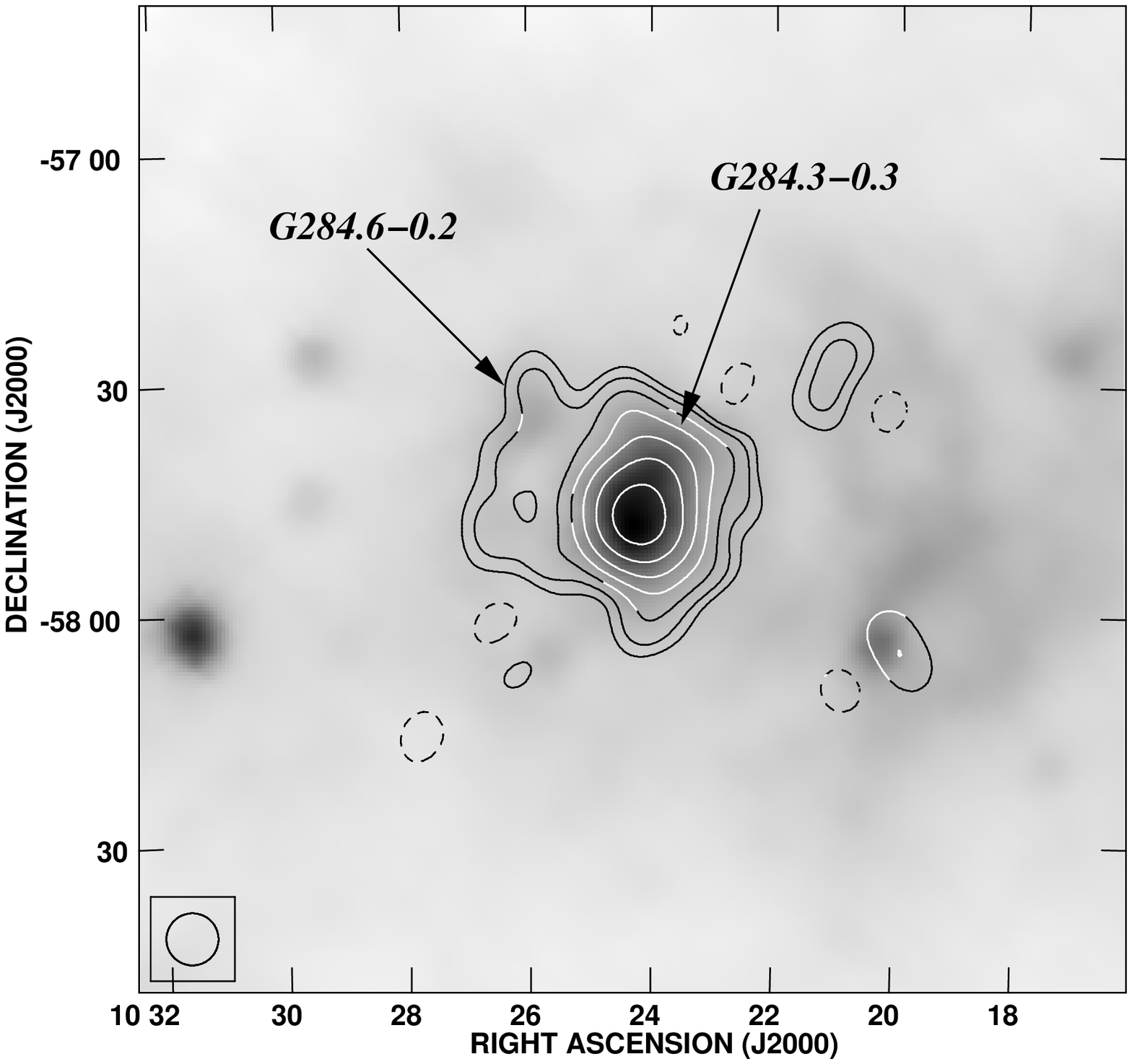}
 \caption{Map of the $G284.3-0.3$ (RCW49) region. CBI 31~GHz contours are overlaid on a greyscale image of the {\it IRAS} $100~\mu$m map, with a square-root stretch. Contours are at $-1~(dashed),1,2,4,8,16,32$ and 64 per cent of the peak intensity, $S_{p}=79.6$~Jy~beam$^{-1}$. The uniform-weighted beam is $6.78 \times 6.78$ arcmin. \label{fig:g284_cbi_map}}
\end{center}
\end{figure}

\begin{figure}
\begin{center}
\includegraphics[width=0.48\textwidth,angle=0]{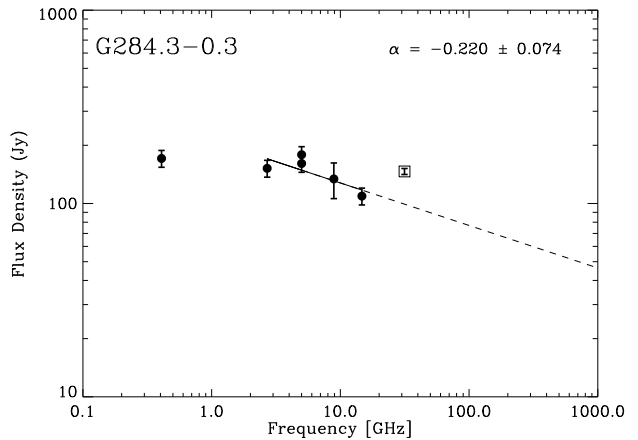}
 \caption{Spectrum of $G284.3-0.3$ (RCW49). Data points ({\it solid circles}) are integrated flux densities taken from the literature (see text). Uncertainties of 10 per cent were assumed when no error was given. The CBI 31~GHz value is plotted as a {\it square} symbol. The best fitting power-law ({\it solid line}) was fitted to the data over the range $2.7-15$~GHz and extended to higher frequencies ({\it dashed line}).}
\label{fig:g284_spec}
\end{center}
\end{figure}

The bright HII region $G284.3-0.3$ (RCW49, NGC3247, MSH10-54, Gum29;
RA(J2000)$=10^{h}24^{m}15^{s}$, Dec.(J2000)$=-57^{d}46^{m}58^{s}$ )
has a peak flux density of 79.6 Jy~beam$^{-1}$ in the 31~GHz CBI
primary-beam-corrected map shown in Fig.~\ref{fig:g284_cbi_map}. The
synthesised beam has a FWHM of $6.78$~arcmin. There are low level
extensions to the north and east that include the diffuse source
$G284.6-0.2$. Wilson et al.~(1970) note that the shoulder of emission
to the east is probably not related to the brighter object. The CBI
map agrees very well with the 5 GHz map \cite{Goss70} and the
100~$\mu$m map (Fig.~\ref{fig:g284_cbi_map}). A single Gaussian fit to
the brightest object (not including the NE extension, but allowing for
a curved baseline) gave an integrated flux density of $146.5 \pm
5.2$~Jy with a deconvolved size of $7.8 \times 5.6$~arcmin. The
eastern and northern extensions contain integrated flux densities of
$\approx 25$~Jy and $\approx 30$~Jy, respectively. However, they have
negligible effect ($\approx 1$~per cent) on the fitting of the much
brighter and compact component, $G284.3-0.3$. Stellar emission from
massive O-type stars, such as those found in Westerlund 2 cluster, is
negligible. The strongest emission is likely to
come from colliding winds in Wolf-Rayet systems that is typically at the
mJy level \cite{Benaglia05,Rauw07}.

The spectrum of $G284.3-0.3$ is plotted in
Fig.~\ref{fig:g284_spec}. Data points from the literature are as for
$G267.9-1.1$ where data are available, with the addition of 5.0~GHz
\cite{Caswell87}. The 5~GHz value appears to be above the line of the
other data points with $S_i=178.8$~Jy. We note that Caswell \& Haynes
(1987) find a flux density of 161~Jy at 5~GHz, but for a slightly
smaller size of FWHM $7\times5$~arcmin, that is more consistent with
the fitted spectrum. We also performed a re-analysis of the Parkes
6~cm data \cite{Haynes78} and find $S_i=175$~Jy. Still, it is possible
that the 14.7~GHz is a little low due to a smaller beam and smaller
fitted area of $5.0\times5.2$~arcmin.  A power-law fit over the range
$2.7-15$~GHz has a spectral index $\alpha=-0.220\pm0.074$, and the CBI
point is well above this line; the predicted 31~GHz flux density is
$99.6\pm 13.4$~Jy. As can be seen from the spectrum in
Fig.~\ref{fig:g284_spec}, the CBI point appears to be significantly
above the expected emission from optically thin free-free alone. For
this model, the excess is $46.9\pm14.4$~Jy. This is a detection of
excess emission at the $3.3\sigma$ level and could account for 32 per
cent of the total emission at 31~GHz. The significance increases
further when fixing the spectral index to the slightly flatter value
of $-0.12$, which is more consistent with that expected from theory
for $T_e\approx8500$K \cite{Azcarate92}. For this model, the excess is
$29.5\pm7.8$~Jy ($3.8\sigma$). Only when omitting the 14.7~GHz data
point did the CBI point come in line with the model with a spectral
index $\alpha=+0.004\pm0.059$. 

Using only the 2.7~GHz and 8.9~GHz data
gave a spectral index of $-0.11 \pm 0.13$, consistent with the 408~MHz
data point. In this case, there still remained a significant
($2.5\sigma$) excess at 31~GHz of $29.1 \pm 11.8$~Jy. The spectral
index within the CBI band is $\alpha=-0.11\pm0.09$.  Cross-correlation
of the simulated 5~GHz visibilities and CBI visibilities show a very
tight correlation of $P=0.94$ with a mean slope of
$0.02387$K~K$^{-1}$, which corresponds to $\alpha=-0.05\pm0.10$. We
therefore consider this a tentative detection of anomalous
emission. Clearly, more precise data in the $5-15$~GHz range are
required to determine the free-free spectrum more accurately and
confirm this result.

The 100~$\mu$m image is well-matched to the CBI image. A simulated
observation, assuming a dust emissivity of $10~\mu$K~(MJy/sr)$^{-1}$
gave a flux density of $34.5 \pm 1.0$~Jy; the error was estimated from
trying different fitting boxes. The 31~GHz excess seen in $G284.3-0.3$
therefore has a 100~$\mu$m emissivity of $13.6\pm
4.2~\mu$K~(MJy/sr)$^{-1}$. For a fixed spectral index, $\alpha=-0.12$,
the emissivity becomes $8.6\pm 2.3~\mu$K~(MJy/sr)$^{-1}$.

%%%%%%%%%%%%%%%%%%%%%%%%%%%%%%%%%%%%%%%%%%%%%%%%%%%%%%%%%%%%%%%%%%

\subsection{$G287.4-0.6$ (Carina nebula) region}

\begin{figure}
\begin{center}
\includegraphics[width=0.48\textwidth,angle=0]{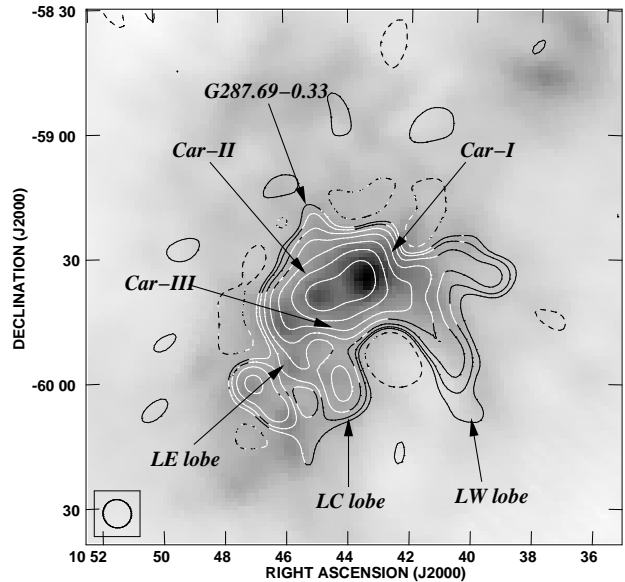}
 \caption{Map of the $G287.9-1.1$ (Carina nebula) region. CBI 31~GHz contours are overlaid on a greyscale image of the {\it IRAS} $100~\mu$m map, with a square-root stretch. Contours are at $-1~(dashed),1,2,4,8,16,32$ and 64 per cent of the peak intensity, $S_{p}=45.8$~Jy~beam$^{-1}$. The uniform-weighted beam is $6.9 \times 6.7$ arcmin. \label{fig:g287_cbi_map}}
\end{center}
\end{figure}

\begin{figure}
\begin{center}
\includegraphics[width=0.48\textwidth,angle=0]{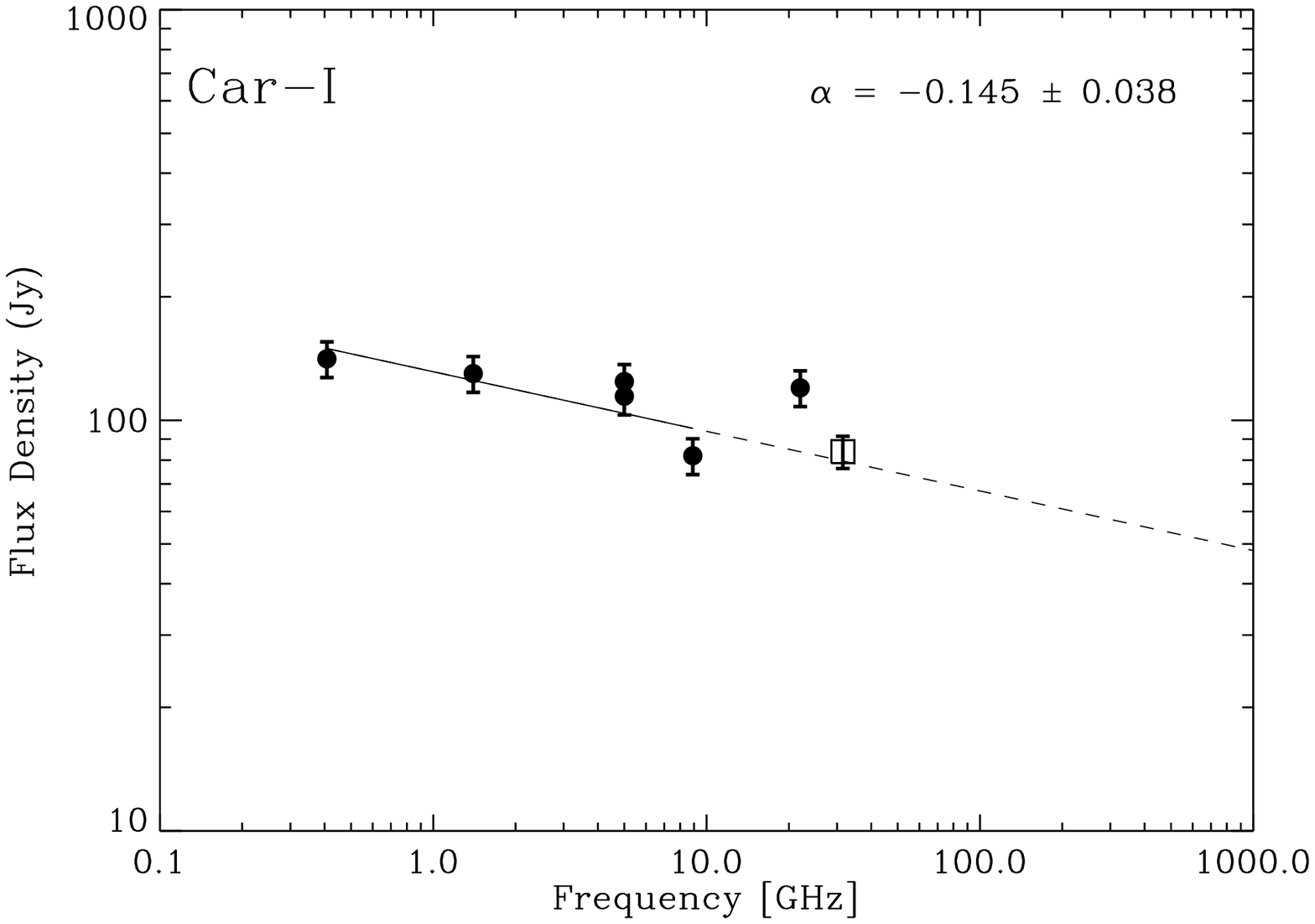}
 \caption{Spectrum of Car-I. Data points ({\it solid circles}) are integrated flux densities taken from the literature (see text). Uncertainties of 10 per cent were assumed when no error was given. The CBI 31~GHz value is plotted as a {\it square} symbol. The best fitting power-law ({\it solid line}) was fitted to the data over the range $0.4-9$~GHz and extended to higher frequencies ({\it dashed line}).}
\label{fig:car-I_spec}
\end{center}
\end{figure}

\begin{figure}
\begin{center}
\includegraphics[width=0.48\textwidth,angle=0]{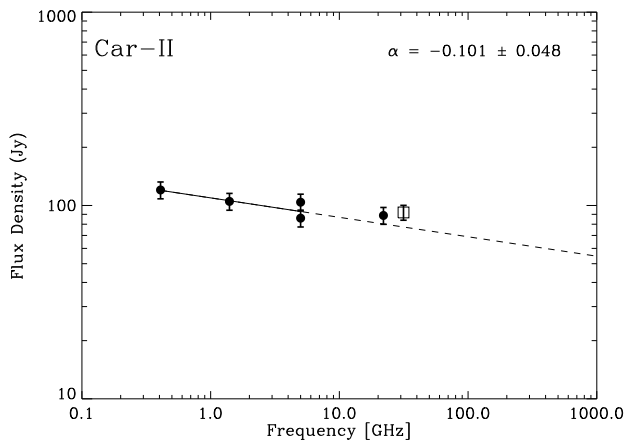}
 \caption{Spectrum of Car-II. Data points ({\it solid circles}) are integrated flux densities taken from the literature (see text). Uncertainties of 10 per cent were assumed when no error was given. The CBI 31~GHz value is plotted as a {\it square} symbol. The best fitting power-law ({\it solid line}) was fitted to the data over the range $0.4-5$~GHz and extended to higher frequencies ({\it dashed line}).}
\label{fig:car-II_spec}
\end{center}
\end{figure}

The Carina nebula (RCW53,NGC3372,MSH10-57,Gum 33, Keyhole nebula)
consists of two main radio sources: Car-I (NW) ($G287.4-0.6$) and
Car-II (SE) ($G287.6-0.6$). These are excited by the young open clusters Tr14
and Tr16 \cite{Tateyama91} and are at a common distance of $2.2 \pm
0.2$~kpc. A number of weaker sources have been identified within the
Carina nebula complex, which covers an area of 4 sq. degrees.

The CBI 31~GHz primary-beam-corrected map is shown in
Fig.~\ref{fig:g287_cbi_map}, with a synthesised beam
$6.9\times6.7$~arcmin and peak flux density,
$S_p=45.8$~Jy~beam$^{-1}$. There is much extended emission in this
region, particularly the several ``lobes'' that extend to the south,
which are also seen at lower frequencies \cite{Whiteoak94,Duncan95}
and are thought to be non-thermal \cite{Tateyama91}.  The non-thermal
lobes (LE, LC and LW) of Tateyama et al.~(1991) are clearly seen as
extensions to the south of Car-I/Car-II in
Fig.~\ref{fig:g287_cbi_map}, including Car-III (southern lobe of
Car-II). The source $G287.69-0.33$ can be identified $\sim 20$~arcmin
to the north of Car-II with a peak flux density of $\approx 2$~Jy. At
this resolution, the brighter central region can just be resolved into
the two known components, Car-I and Car-II, which are clearly seen in
the 100~$\mu$m map (Fig.~\ref{fig:g287_cbi_map}).

Two Gaussians were fitted simultaneously to the central part of the
primary-beam-corrected image with a baseline slope to account for the
surrounding extended emission. We found that two Gaussians could be
well-fitted to the data with $S_{i}^{31}=83.9\pm7.6$~Jy and
$S_{i}^{31}=92.1\pm8.1$~Jy, for Car-I and Car-II, respectively
(Table~\ref{tab:flux}). The larger errors reflect the fact that the
components are slightly confused at this resolution. Their deconvolved
sizes were measured to be $8.8\times6.1$~arcmin and
$9.4\times6.9$~arcmin, respectively.

The spectrum of Car-I is shown in Fig.~\ref{fig:car-I_spec}. Data
points from the literature are as for $G267.9-1.1$ where data are
available, with the addition of  1.4~GHz \cite{Retallack83}, 8.9~GHz
\cite{Huchtmeier75} and 22~GHz \cite{Tateyama91}. We re-analysed the
Parkes 5~GHz map of Haynes et al.~(1978) and found it to be consistent
with the Goss \& Shaver (1970) result. In Fig.~\ref{fig:car-I_spec} we
include the 22~GHz flux density from Tateyama et al.~(1991) by scaling
their peak flux density with their reported source size, but do not
include it in the fit due to possible errors in this extrapolation. It
is interesting to see the 22~GHz data point is well above the
free-free model. This could be real and is consistent with spinning
dust models that predict a peak at this frequency (Draine \& Lazarian
1998a,b). The spectrum lies close to the optically thin free-free
value down to 408~MHz \cite{Gardner70}. The best-fitting power-law
over the range $0.4-9$~GHz has a spectral index
$\alpha=-0.145\pm0.038$, consistent with that predicted by theory for
$T_e\approx 6600-7400$~K \cite{Gardner70,Caswell87}. The best-fitting
spectral index within the CBI band is $\alpha=-0.13\pm0.09$. The model
predicts a 31~GHz flux density $S_{i}^{31}=79.8\pm8.3$~Jy and the CBI
31~GHz data point fits well within this model with an upper limit for
an excess of 24.8~Jy (95 per cent c.l.). For a fixed spectral index,
$\alpha=-0.12$, the predicted 31~GHz flux was $84.8\pm3.8$~Jy
corresponding to an upper limit of $<16.0$~Jy.

The CBI-simulated 100~$\mu$m map, scaled with
$10~\mu$K~(MJy/sr)$^{-1}$ gives a flux density of $28.0\pm 2.0$~Jy for
a point-like source. This translates to an upper limit on the excess
dust emissivity of $<6.1~\mu$K~(MJy/sr)$^{-1}$ at the 95 per cent
c.l. For the fixed spectral index model, the emissivity is
$<5.7~\mu$K~(MJy/sr)$^{-1}$.

The spectrum of Car-II is shown in Fig.~\ref{fig:car-II_spec} with the
same data plotted as for Car-I, except for omitting the 8.9~GHz value
from Huchtmeier \& Day (1975), which appears to be anomalously low. This is probably due
to a mismatch in fitted source size and proximity of Car-I. The
free-free emission remains optically thin down to 408~MHz
\cite{Gardner70} with a spectral index $\alpha=-0.101\pm0.048$ fitted
over the range $0.4-5$~GHz, again close to the theoretical value. The
predicted 31~GHz flux density is then $77.5\pm11.1$~Jy. The CBI data
point lies slightly above the prediction with an upper limit of
$<38.1$~Jy (95 per cent c.l.) for excess emission. Although not a
statistically strong detection, it is interesting to see that the
22~GHz value is also above the free-free model alongside the 31~GHz
data point. For a fixed spectral index, $\alpha=-0.12$, the predicted
flux density is $73.4\pm3.7$~Jy, allowing an excess of $18.7 \pm
8.9$~Jy {\it i.e.} a $2.1\sigma$ detection.

The CBI-simulated 100~$\mu$m map, scaled with
$10~\mu$K~(MJy/sr)$^{-1}$ gives a flux density of $24.0\pm 1.6$~Jy for
a point-like source. This translates to an upper limit on the excess
dust emissivity of $<15.9~\mu$K~(MJy/sr)$^{-1}$. For the fixed
spectral index model, the emissivity is at
$7.8\pm3.7~\mu$K~(MJy/sr)$^{-1}$, as summarised in
Table~\ref{tab:flux}.

%%%%%%%%%%%%%%%%%%%%%%%%%%%%%%%%%%%%%%%%%%%%%%%%%%%%%%%%%%%%%%%%%%

\subsection{$G291.6-0.5/G291.3-0.7$ (RCW57) region}

\begin{figure}
\begin{center}
\includegraphics[width=0.48\textwidth,angle=0]{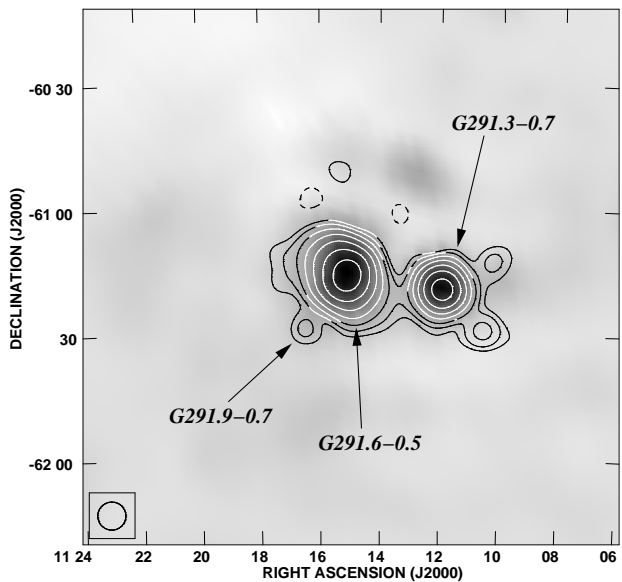}
 \caption{Map of the $G291.6-0.5$ (NGC3603) and $G291.7-0.3$ (NGC3576) region. CBI 31~GHz contours are overlaid on a greyscale image of the {\it IRAS} $100~\mu$m map, with a square-root stretch. Contours are at $-1~(dashed),1,2,4,8,16,32$ and 64 per cent of the peak intensity, $S_{p}=88.0$~Jy~beam$^{-1}$. The uniform-weighted beam is $6.7 \times 6.7$ arcmin. \label{fig:g291_cbi_map}}
\end{center}
\end{figure}

\begin{figure}
\begin{center}
\includegraphics[width=0.48\textwidth,angle=0]{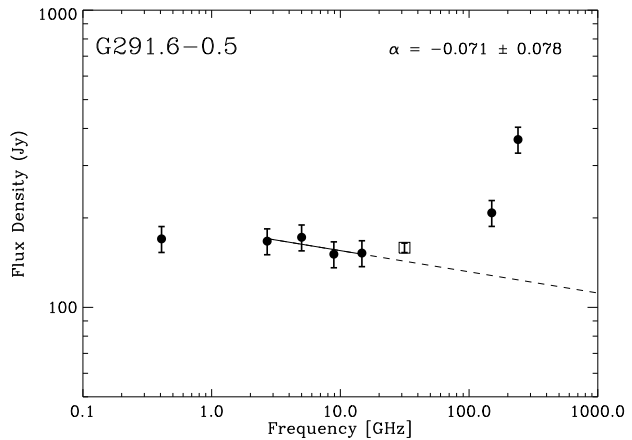}
 \caption{Spectrum of $G291.6-0.5$. Data points ({\it solid circles}) are integrated flux densities taken from the literature (see text). Uncertainties of 10 per cent were assumed when no error was given. The CBI 31~GHz value is plotted as a {\it square} symbol. The best fitting power-law ({\it solid line}) was fitted to the data over the range $5-15$~GHz, fixing the spectral index $\alpha=-0.12$ (see text) and extended to higher frequencies ({\it dashed line}). \label{fig:g291.6_spec}}
\end{center}
\end{figure}

\begin{figure}
\begin{center}
\includegraphics[width=0.48\textwidth,angle=0]{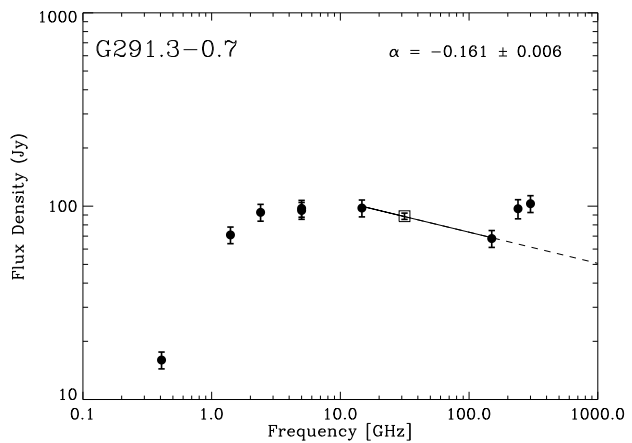}
 \caption{Spectrum of $G291.3-0.7$. Data points ({\it solid circles}) are integrated flux densities taken from the literature (see text). Uncertainties of 10 per cent were assumed when no error was given. The CBI 31~GHz value is plotted as a {\it square} symbol. The best fitting power-law ({\it solid line}) was fitted to the data over the range $15-150$~GHz including the CBI data point (see text) and extended to higher frequencies ({\it dashed line}). \label{fig:g291.3_spec}}
\end{center}
\end{figure}

The RCW57 region is dominated by two bright \hii~regions: $G291.6-0.5$
(NGC3603) and $G291.3-0.7$ (NGC3576), which are two of the highest
luminosity optically visible \hii~regions in the Galaxy
\cite{Goss69}. The 31~GHz CBI map, with a synthesised beam FWHM of
$6.7$~arcmin, is shown in Fig.~\ref{fig:g291_cbi_map}. The two
\hii~regions dominate the map: $G291.6-0.5$ is the larger eastern
component at the center of the image (RA(J2000)$=11^h15^m00^s$,
Dec(J2000)$=-61^d16^m00^s$; $G291.3-0.7$ is the more compact component
located $\approx 25$~arcmin to the west. The CBI 31~GHz map shows
NGC3603 as the brighter and slightly extended source with peak flux
density $S_p= 88.0$~Jy~beam$^{-1}$ and NGC3576 has
$S_p=79.8$~Jy~beam$^{-1}$ after correcting for the primary beam. The
31~GHz map agrees very well with the low frequency GHz maps
\cite{Shaver70a,Goss70} and the 100~$\mu$m map
(Fig.~\ref{fig:g291_cbi_map}). Some low-level extended emission is
also detected in the vicinity of the dominant \hii~regions. A compact
source to the south-east of NGC3603 is detected and is identified as
$G291.9-0.7$ with an integrated flux density of $\approx 2$~Jy.

Single Gaussians were fitted to the 2 bright sources in the CBI
primary-beam-corrected map. For $G291.6-0.5$, we find an integrated
flux density $S_i=158.7 \pm 5.8$~Jy with a deconvolved size $7.1
\times 7.1$ arcmin. The spectrum is plotted in
Fig.~\ref{fig:g291.6_spec}. Data points from the literature are as for
$G267.9-1.1$ where data are available, with the addition of 150 and
240~GHz data (Sabattini et al.~2005). The spectrum remains relatively
flat up to a frequency of several GHz possibly indicating optically
thick components. Nevertheless, we fitted a power-law to data in the
range $2.7-15$~GHz and obtained $\alpha=-0.071 \pm 0.078$. This model
gave a predicted 31~GHz flux of $143.2 \pm 19.0$ or an upper limit to
an excess component of $<50.3$~Jy (95 per cent c.l.).  For a fixed
spectral index, $\alpha=-0.12$, the prediction becomes
$132.5\pm6.6$~Jy, or an excess of $26.2\pm 8.8$~Jy {\it i.e.} a
$3\sigma$ detection. The spectral index within the CBI band was found
to be $\alpha=-0.12\pm0.09$, consistent with a typical electron
temperature of $T_{e}\approx 7000-8000$~K
\cite{Wilson70,McGee75,McGee81,dePree99}. The two data points at 150
and 240~GHz (Sabattini et al.~2005) provide a useful limit to the
contribution of vibrational (thermal) dust emission, where they find a
dust temperature $T_{d}=25.6$~K. Unless there exists a very cold dust
component, the contribution from vibrational dust at 31~GHz is
relatively small. Extrapolating from the Sabattini et al. values,
assuming an emissivity index $\beta=+2.0$, gives a flux density of
8.9~Jy, or 6 per cent of the total; this may explain the small excess
observed and slightly flatter spectral index. Indeed, making a
correction for the vibrational dust component, reduces the
significance of the  detection (for a fixed spectral index) to
$1.2\sigma$.

The CBI-simulated $100~\mu$m, scaled with $10~\mu$K~(MJy/sr)$^{-1}$
gives a peak brightness of 12.1~Jy~beam$^{-1}$ and an integrated flux
density $S_{i}^{31}=32.1\pm3.0$~Jy. The 95 per cent c.l. upper limit
on the excess dust emissivity is then $15.7~\mu$K~(MJy/sr)$^{-1}$. For
the fixed spectral index model, with no correction for a vibrational
dust contribution, the emissivity is $12.3\pm
4.3~\mu$K~(MJy/sr)$^{-1}$, as summarised in Table~\ref{tab:flux}.

For $G291.3-0.7$, we find $S_i=88.8\pm3.3$~Jy with a deconvolved size
$2.6\times2.6$~arcmin. The spectrum is plotted in
Fig.~\ref{fig:g291.3_spec} with data taken from the literature. The
turn-over is much more gradual indicating optically thick components
and only becomes truly optically thin above $\sim 10$~GHz. With so few
data points to fit, we included the 14.7~GHz, 150~GHz and CBI 31~GHz
data point itself in the fit. The best-fitting spectral index in the
range $15-150$~GHz is $\alpha=-0.161 \pm 0.006$ and is an excellent
fit to the data. The upper limit is 6.7~Jy (95 per cent c.l.) for an
additional component. From the data at 240 and 300~GHz, there appears
to be a small contribution from vibrational dust at 150~GHz, which was
found to be typically warmer ($T_d=31.3$~K) than for $G291.6-0.5$
(Sabattini et al.~2005). This is somewhat discrepant with the values
found by Kuiper et al.~(1987) who find warmer dust temperatures,
$T_{d}\approx 50$~K, based on the $60/100~\mu$m ratios. Extrapolating
from the Sabattini et al. values, for $\beta=+2.0$, gives 1.6~Jy at
31~GHz ($\approx 2$~per cent). This would steepen the spectral index
further and therefore leave the possibility for a small excess at
31~GHz. However, we tried several different fits with varying
assumptions (e.g. including lower frequency data), which did not allow
a significant additional component at 31~GHz.

The CBI-simulated $100~\mu$m, scaled with $10~\mu$K~(MJy/sr)$^{-1}$
has a peak brightness of 5.5~Jy~beam$^{-1}$ and in integrated flux
density $S_{i}^{31}=9.8\pm0.9$~Jy. The upper limit on the excess dust
emissivity is then $<6.7~\mu$K~(MJy/sr)$^{-1}$ when the fit was done
including the CBI data point. This will under-estimate any possible
excess emission since including the CBI point reduces the allowable
range. However, it is such a good fit to the model, the fit is
unlikely to change by much. For a fixed spectral index,
$\alpha=-0.12$, the upper limit becomes 15.8~Jy (95 per cent c.l.), or
a dust emissivity $<16.1~\mu$K~(MJy/sr)$^{-1}$. As with $G291.6-0.5$,
the data points at 150 and 240~GHz (Sabattini et al.~2005) suggest
there may be a contribution from vibrational dust emission at 31~GHz
of a few Jy, or a few per cent of the total 31~GHz flux density and a
much larger fraction at 150~GHz. Without more data points, and
detailed modelling of the dust spectrum, it is difficult to calculate a
more precise limit for this source. However, our upper limits can be
considered as being conservative since these corrections are likely to
steepen the free-free model allowing more room for excess
emission. Nevertheless, it is clear from Fig.~\ref{fig:g291.3_spec}
that the CBI 31~GHz data point fits in well with the other data points
following a smooth curve, leaving little or no room for possible
excess emission.

\subsection{Polarisation limits}
\label{sec:polarisation}

Stokes $Q$ and $U$ maps were made for each region and imaged/cleaned
using the same procedure as for the total-intensity maps. Polarised
intensity maps, $P=(\sqrt{Q^2+U^2}-C)$, were made using the {\sc aips}
task {\sc comb}, where $C$ is the correction for the Ricean noise
bias, using estimates of the noise from areas of the map away from the
primary beam. For all four regions, small polarised signals were
detected.  For both $G267.9-1.1$ and $G284.3-0.3$, a ring-like
structure was observed with a peak frequency centred close to the map
centre. In the Carina nebula map, we observed two point-like peaks
centred on Car-I and Car-II. For $G291.6-0.5$ a similar faint
ring-like feature is seen. The largest polarised signal, at
480~mJy~beam$^{-1}$ or 0.61 per cent polarisation fraction, was
observed in $G291.3-0.7$; a highly significant ($>10\sigma$) detection
is observed in both $Q$ and $U$ maps. The peak polarised flux
densities for all the \h2 regions are given in
Table~\ref{tab:polarisation} along with the polarised fraction
calculated from the ratio of peak flux densities.

\begin{table}
\caption{31~GHz polarised intensity measurements. Statistically significant polarisation fractions were detected in all the sources at similar level, but are likely to be contaminated by leakage terms and hence are considered upper limits (see text).}
\begin{tabular}{lccc} \hline
Source   &$S_{p}^{31}$  &r.m.s. noise  &Polarisation   \\
         &(mJy~bm$^{-1}$)    &(mJy~bm$^{-1}$)    &fraction (per cent)\\     
\hline
$G267.9-1.1$ &348       &75            &$0.28\pm0.06$ \\

$G284.3-0.3$ &190       &25            &$0.24\pm0.03$ \\

Car-I        &119       &23            &$0.32\pm0.06$ \\

Car-II       &123       &23            &$0.33\pm0.06$ \\

$G291.6-0.5$ &204       &35            &$0.25\pm0.04$ \\

$G291.3-0.7$ &480       &35            &$0.61\pm0.04$ \\
\hline
\end{tabular}
\label{tab:polarisation}
\end{table}

Given the brightness in total intensity, the observed polarised
signals are unlikely to be real since no corrections were made for
instrumental leakage terms, which are expected to be at the $\sim 1$
per cent level. The observed polarisation is therefore consistent with
polarisation generated by the instrument itself, which is discussed
further in section \ref{sec:pol_limits}.

%This would naturally explain the similarity of
%observed polarisation fractions, which were all at 0.3 per cent,
%except for $G291.3-0.7$, which was at 0.61 per cent. This may be due to
%the fact that $G291.3-0.7$ is located away from the map centre by
%almost half a primary beam width. This could contribute extra leakage
%and/or errors due to the primary beam correction. We therefore take
%these values to be upper limits to the polarisation on these angular
%scales. This is consistent with little or no polarisation expected for
%pure free-free emission. 

%Moreover, this verifies that the CBI
%instrument has good polarisation properties in that the instrumental
%leakage is below 1 per cent, as recently claimed for detections of CMB
%polarisation \cite{Readhead04b,Sievers07}.

%%%%%%%%%%%%%%%%%%%%%%%%%%%%%%%%%%%%%%%%%%%%%%%%%%%%%%%%%%%%%%%%%%
%%%%%%%%%%%%%%%%%%%%%%%%%%%%%%%%%%%%%%%%%%%%%%%%%%%%%%%%%%%%%%%%%%

\section{Discussion}

\subsection{Free-free emission}

\begin{table*}
\caption{Physical and observed properties. Dust temperatures are taken from Kuiper et al.~(1987). The average dust temperature at high Galactic latitudes is 18.2~K (Schlegel et al.~1998). Electron temperatures were average values taken from the literature (see text for details).}
\begin{tabular}{lccccc} \hline
Source &Radio size        &FIR size              &$T_{d}$&$\alpha$ &$T_{e}$ \\
       &$@31$~GHz (arcmin) &$@100~\mu$m (arcmin) &(K)    &($S\propto \nu^{\alpha}$) &(K) \\
\hline
$G267.9-1.1$ &$2.6\times2.2$&$4.9\times4.2$      &45    &$-0.115\pm0.023$  &7500 \\ 

$G284.3-0.3$ &$7.8\times5.6$&$7.9\times4.9$      &50    &$-0.220\pm0.074$  &8500 \\ 

Car-I        &$8.8\times6.1$&$6.0\times5.2$      &48    &$-0.145\pm0.038$  &7000 \\ 

Car-II       &$9.4\times6.9$&$9.7\times5.7$      &70    &$-0.101\pm0.048$  &6600 \\ 

$G291.6-0.5$ &$7.1\times7.1$&$11.6\times7.1$     &55    &$-0.071\pm0.078$  &7500 \\ 

$G291.3-0.7$ &$2.6\times2.6$&$4.9\times4.7$      &45    &$-0.161\pm0.006$  &7500 \\ 
\hline
\end{tabular}
\label{tab:properties}
\end{table*}

All the \hii~regions discussed in this paper are dominated by thermal
free-free emission, which becomes optically thin at frequencies
$\gtsim 1$~GHz. These bright sources usually consist of many compact
objects that are unresolved by the CBI beam, many of which contain
substantial dust which emits primarily in the FIR band ($\sim
100~\mu$m) within a similar volume (Table~\ref{tab:properties}). At
higher frequencies, the blackbody tail from the vibrating dust
mechanism typically dominates and can extend down to frequencies $\sim
100$~GHz. These two emission mechanisms largely explain the general
shape of the spectrum over a wide range of frequencies, from the radio
to the mid-IR. Indeed, we have found that the 31~GHz flux observed
with the CBI is broadly consistent with free-free emission when
combined with multi-frequency data taken from the literature.

The original purpose of this study was to search for evidence of
spinning dust, which would show up as an additional excess component
at 31~GHz, which is close to the peak of current models of spinning
dust (Draine \& Lazarian 1998a,b). There are inconsistencies with some
of the data in the literature, where calibration errors are typically
10 per cent on quoted flux densities. In addition to this, there can
be difficulties when comparing data taken at different resolutions and
where different fitting techniques have been employed. Fortunately,
the spectrum of free-free emission is well understood. When the
radiation becomes optically thin it has a well-defined spectral index
that varies slowly with frequency and electron temperature. In fact,
many authors simply fix the spectral index to the canonical radio
spectral index, $\alpha=-0.1$. At higher frequencies ($\sim 30$~GHz),
it steepens slightly to $\alpha=-0.14$ for $T_e=7000-8000$~K
\cite{Dickinson03}.

We have found that the best-fitting spectral index (Table
\ref{tab:properties}), not including CBI data, for all \hii~regions
was essentially consistent with this range of values. The electron
temperatures for all the \hii~regions are within the range $\sim
7000-8000$~K (Table~\ref{tab:properties}). Furthermore, the CBI 31~GHz
data point, was found to be close to the predicted flux density from a
simple power-law fit to data from the literature. This confirms the
dominance of free-free emission in bright \hii~regions. Fits were also
made with a fixed spectral index, $\alpha=-0.12$, which is the mean
spectral index expected for free-free emission for the range
$1-30$~GHz  for $T_e=8000$~K \cite{Dickinson03}. Although this
artificially reduces the error in the model, it can help limit the
impact of low or high data points that can artificially bias the
spectral index, particularly when only a few data points are being
fitted. For most sources, we found that the results remained stable
either way.

\subsection{Anomalous dust emission}

We have found that at least one of the sources, $G284.3-0.3$ (RCW49),
shows evidence for a significant excess component, suggestive of
spinning dust. Furthermore, it is compelling that all six sources are
found to have a slightly higher flux density at 31~GHz than the
predicted value given by a power-law model for the free-free
emission. The average $100~\mu$m dust emissivity for all the 6 sources
is $3.3\pm1.7~\mu$K~(MJy/sr)$^{-1}$, which corresponds to a 95 per
cent upper limit of $6.1~\mu$K~(MJy/sr)$^{-1}$. We have discussed some
possible systematics that may lead to this apparent excess, but given
the conservative error bars assigned to the data, and the
relative accuracy of the CBI data, this result is unlikely to be due
to a systematic error. Moreover, no flux loss correction was made to
the CBI data points, since it was shown to be a small correction for
sources comparable to the beam size (see
section~\ref{sec:imaging}). 

The most significant ($3.3\sigma$) result was for $G284.3-0.3$
(RCW49). We consider this to be a tentative detection. As remarked upon in
section~\ref{sec:g284}, there is some level of inconsistency in the
lower frequency data in the range $5-14$~GHz. In particular, the 5~GHz
data point seems high relative to the other frequencies, yet we
obtained a consistent value when we independently analysed the Parkes
6~cm map. Moreover, the 14~GHz point appears to be on the low side,
while the 9~GHz value has a larger error than its neighbours. For
example, taking just the 5~GHz data and the 31~GHz data alone, the
spectral index is significantly flatter and is more consistent with
free-free alone. Clearly more precise data in the range $5-20$~GHz
data is required to clarify the situation. 

If spinning dust emission is found to be a significant fraction of the
31~GHz flux emission, it would be expected to originate from very
small grains that can spin fast enough to produce observable
emission. The smallest grains, polycyclic aromatic hydrocarbons (PAHs)
are one possibility. PAHs are most readily identified as broad lines
in the mid-IR spectrum that have been seen in many \hii~regions and
PNe. The observed survival of
small dust grains in hostile environments is difficult to reconcile
with models \cite{Spitzer78}, yet strong mid-IR PAH signals
are observed in active star-forming regions including RCW49
\cite{Churchwell04}. The spinning dust mechanism appears therefore to
be still viable in such environments. It is rather surprising that the spectral
index at $\sim 30$~GHz is so similar to the canonical free-free value
($\alpha \approx -0.1$). However, the spinning dust spectrum is
expected to turn over in the range $\sim 20-40$~GHz and hence it may
appear to be locally flat in this range.

%We also note that the dust temperature,
%derived from the $60/100~\mu$m ratios (Table~\ref{tab:properties}),
%was found to be relatively high for this source, at $T=50$~K
%\cite{Kuiper87}. Average dust temperatures of \hii~regions are
%typically in the range $T_{d}=30-50$~K, although Car-II with
%$T_d=70$~K is an exceptional case \cite{Kuiper87}.

\subsection{Anomalous dust emissivity}

The radio emission from dust in HII regions is found to be
a factor of 3-4 less than in the cooler diffuse dust at
intermediate latitudes. The limits on excess emission at 31~GHz have been converted to a dust
emissivity, relative to the IRIS re-processed version of the {\it IRAS}
$100~\mu$m map \cite{Miville05}, which has units MJy~sr$^{-1}$, thus our emissivities
have units $\mu$K~(MJy/sr)$^{-1}$. We did this for simplicity and
because it is model independent\footnote{Some authors have calculated
dust emissivities relative other standards, including the DIRBE
$140~\mu$m map, the Finkbeiner, Davis \& Schlegel (1999) model 8 map
normalised at 94~GHz, or the hydrogen column density, $n_{\rm H}$,
estimated from the $100~\mu$m map of Schlegel, Finkbeiner \& Davis
(1998); see Finkbeiner (2004) for a useful discussion of units.}. From
CMB data at frequencies $\sim 31$~GHz, and at high Galactic latitudes,
the dust emissivity has a typical value of $\sim
10~\mu$K~(MJy/sr)$^{-1}$, with variations of about a factor of $\sim
2$ \cite{Davies06}. We can immediately see that our tentative
detection, in $G284.3-0.3$, is consistent with the dust emissivity
found at high latitude. On the other hand,  the upper limits listed in
Table~\ref{tab:flux} indicate that the dust emissivity is considerably
lower than that found at high latitudes; the average emissivity for
all 6 \hii~regions (when the spectral index was fitted for) is
$3.3\pm1.7~\mu$K~(MJy/sr)$^{-1}$. In other words, if the spinning dust
were to emit at the same levels as seen at in more quiescent high
latitude regions of sky (at least relative to the $100~\mu$m intensity
map), we would have detected a larger excess in most of the sources
studied here.

In Table~\ref{tab:emissivities}, we have listed the 31~GHz normalised
dust emissivities for \hii~regions and the cooler dust clouds from the
literature. This emphasises that the dust emissivity appears to be
less in the \hii~regions than in the diffuse interstellar medium by a
factor of $\sim 3-4$, but where the average dust temperature is
$18.2$~K (Schlegel et al.~1998). It is also clear that the $T^{1.6}$
scaling of emissivity at high latitudes, found by Davies et
al.~(2006), does not hold in these regions; the warmer dust does not
emit at higher levels relative to $100~\mu$m data. This is presumably
due to the different conditions in the interstellar medium, where in
\hii~regions there is a considerably larger flux of $UV$-photons from
O-B stars that formed the ionised regions, which in turn disassociates
the smaller grains required for current models of spinning dust
emission. 

H{\sc ii} regions exhibit a wide range of environmental conditions (UV
radiation field, X-rays, $\gamma$-rays, electron temperatures)
which affect the distribution of grain sizes and properties. The
emissivity of spinning dust could then vary considerably from cloud to
cloud \cite{Davies06}. This could explain the apparent lack of anomalous
emission from some regions but not others. It is also possible that
some other mechanism is responsible for the bulk of the anomalous
signal, including magneto-dipole emission \cite{Draine99}, which
strongly depends on the abundance of ferromagnetic material.

\begin{table}
\caption{Comparison of $100~\mu$m dust emissivities for \hii~regions and cooler dust clouds, from data at or near 30~GHz. Data are the mean of the six \hii~regions studied in this paper, LPH96 (Dickinson et al.~2006), average of 15 high latitude regions from {\it WMAP} and the all-sky {\it WMAP} value outside the Kp2 mask (Davies et al.~2006), LDN1622 (Casassus et al.~2006) and $G159.6-18.5$ in the Perseus molecular cloud (Watson et al.~2005). Emissivities, in units $\mu$K~(MJy/sr)$^{-1}$, have been normalised to 31~GHz.}
\begin{tabular}{lcl} \hline
Source   &Dust emissivity         &Reference \\
         &$\mu$K~(MJy/sr)$^{-1}$  &       \\
\hline
{\bf \hii~regions}    &             &       \\
6 \hii~regions (mean) &$3.3\pm1.7$  &This paper.       \\
\vspace{1mm}
LPH96                 &$5.8 \pm 2.3$&Dickinson et al. (2006) \\

{\bf Cool dust clouds} &             &       \\

15 regions {\it WMAP} &$11.2\pm1.5$ &Davies et al. (2006) \\

All-sky {\it WMAP}    &$10.9\pm1.1$ &Davies et al. (2006) \\

LDN1622               &$24.1\pm0.7$ &Casassus et al. (2006) \\

$G159.618.5$          &$17.8\pm0.3$ &Watson et al. (2005) \\
\hline
\end{tabular}
\label{tab:emissivities}
\end{table}

\subsection{Polarisation limits}
\label{sec:pol_limits}

The CBI polarisation maps all showed some polarised emission for these
sources, but at a very low level of 0.3 per cent, except for
$G291.3-0.7$, which was at 0.6 per cent. This may be due to
the fact that $G291.3-0.7$ is located away from the map centre by
almost half a primary beam width. This could contribute extra leakage
and/or errors due to the primary beam correction. We therefore take
these values to be upper limits to the polarisation on these angular
scales. This is consistent with little or no polarisation, as expected for
pure free-free emission. 

Free-free emission is intrinsically unpolarised, but can be polarised
at the edges of \hii~regions by Thomson scattering. The radiation is
then tangentially polarised to the edges of the cloud, at a level that
depends on the viewing angle relative to the incident radiation
\cite{Rybicki79}. At these angular resolutions, we did not expect to
see this effect since the sources are barely resolved in the CBI beam
thus any secondary scattering will be averaged out by the
beam. Spinning dust emission is expected to be weakly polarised at the
few per cent level \cite{Lazarian00}. However, we did expect some
level of instrumental leakage, which converts Stokes $I$ to Stokes $Q$
and $U$, at the level of $\sim 0.5$~per cent based on earlier
observations of W44 \cite{Cartwright05}. We have not attempted to
correct for leakage terms and hence it is not surprising to see such
levels of polarisation. This naturally explains the remarkably similar
polarisation fractions observed in the different \hii~regions. We
therefore consider the quoted polarisation fractions
(Table~\ref{tab:polarisation}) to be upper limits. The level of the
leakage shown here is encouraging since the recent CBI polarisation
results \cite{Readhead04b,Sievers07} had no corrections made for
instrumental leakage. Given the signal-to-noise ratio of the CMB
polarisation detections, this level of leakage can be safely ignored,
and we can be sure that the contamination is certainly below the 1 per
cent level.

For the most significant detection of excess emission in $G284.3-0.3$,
the polarisation limit translates to an upper limit on the spinning
dust polarisation. If indeed 30 per cent of the 31~GHz emission in
$G284.3-0.3$ is anomalous (e.g. from spinning dust), then the
effective upper limit to the polarisation fraction of this component
is $\sim 1$~per cent. This is lower than that observed in the Perseus
cloud, $G159.6-18.5$, which was observed to have a polarisation
fraction of $3.4^{+1.5}_{-1.9}$~per cent \cite{Battistelli06}. Such
low levels of polarisation are consistent with electro-dipole emission
from spinning grains. The slight discrepancy in polarisation level may
be attributed to varying levels of ferromagnetic material, which
through magneto-dipole emission can be much more highly polarised
\cite{Draine99}.

%%%%%%%%%%%%%%%%%%%%%%%%%%%%%%%%%%%%%%%%%%%%%%%%%%%%%%%%%%%%%%%%%%
%%%%%%%%%%%%%%%%%%%%%%%%%%%%%%%%%%%%%%%%%%%%%%%%%%%%%%%%%%%%%%%%%%

\section{Conclusions}

Observations of 6 bright \hii~regions suggest a small amount
of excess emission at 31~GHz, based on fitting a free-free model from data in
the literature. The dominant source of emission comes from optically
thin free-free emission with a spectral index $\alpha \approx
-0.12$. But we find that all the sources were slightly brighter at
31~GHz relative to the simple free-free model.  The average $100~\mu$m
dust emissivity for the 6 \hii~regions was found to be
$3.3\pm1.7~\mu$K~(MJy/sr)$^{-1}$, or a 95 per cent confidence limit of
$<6.1~\mu$K~(MJy/sr)$^{-1}$. This is lower by a factor of $\sim 3-4$
compared to cooler diffuse clouds at high Galactic latitudes
(Table~\ref{tab:emissivities}). However, only one source, $G284.3-0.3$
(RCW49), was found to be statistically significant with a $100~\mu$m
dust emissivity of $13.6\pm 4.2~\mu$K~(MJy/sr)$^{-1}$
($3.3\sigma$). For this source, there are several caveats in
interpreting and using data from the literature, which could reduce
the significance of this result. New data in the range $5-30$~GHz,
particularly from well-calibrated instruments, are required to clarify
whether our tentative detection holds. The dust emissivity, relative
to the $100~\mu$m map, for this object is consistent with that found
in diffuse clouds  at high Galactic latitudes
(Table~\ref{tab:emissivities}). For the majority of the other sources,
only upper limits could be obtained which appear to show that the dust
emissivity is in fact lower than that observed at high Galactic
latitudes.

We observed very low level ($0.3-0.6$ per cent) polarisation at 31~GHz
from all the \hii~regions studied here. The level is consistent with
that expected from instrumental leakage in the CBI instrument. This
validates claims that the instrumental leakage is negligible ($<1$ per
cent) for recent detections of CMB polarisation with the CBI.

%%%%%%%%%%%%%%%%%%%%%%%%%%%%%%%%%%%%%%%%%%%%%%%%%%%%%%%%%%%%%%%%%%
%%%%%%%%%%%%%%%%%%%%%%%%%%%%%%%%%%%%%%%%%%%%%%%%%%%%%%%%%%%%%%%%%%

\section*{ACKNOWLEDGMENTS}

CD thanks Barbara and Stanley Rawn Jr. for funding a fellowship at the
California Institute of Technology for part of this work. We thank the
staff and engineers at the Chajnantor observatory for their hard work
and continuing support. In particular, we thank Cristobal Achermann,
Ricardo Bustos, Rodrigo Reeves and Nolberto Oyarace. SC aknowledges
support from FONDECYT grant 1060827. SC and LB acknowledge support
from the Chilean Center for Astrophysics FONDAP 15010003. Part of the
research described in this paper was carried out at the Jet Propulsion
Laboratory, California Institute of Technology, under a contract with
the National Aeronautics and Space Administration.

%%%%%%%%%%%%%%%%%%%%%%%%%%%%%%%%%%%%%%%%%%%%%%%%%%%%%%%%%%%%%%%%%%
%%%%%%%%%%%%%%%%%%%%%%%%%%%%%%%%%%%%%%%%%%%%%%%%%%%%%%%%%%%%%%%%%%

%\bibliography{cmb_refs}
%\bibliographystyle{mn2e}

%%%%%%%%%%%%%%%%%%%%%%%%%%%%%%%%%%%%%%%%%%%%%%%%%%%%%%%%%%%%%%%%%%%%
%%%%%%%%%%%%%%%%%%%%%%%%%%%%%%%%%%%%%%%%%%%%%%%%%%%%%%%%%%%%%%%%%%%

\bsp % ``This paper has been produced using the ...''

\label{lastpage}

\end{document}